\documentclass[a4paper]{PoS}
\usepackage{colortbl}
\usepackage{mathtools} 
\usepackage{graphicx}
\usepackage{wrapfig}
\usepackage{lineno}
\usepackage{filecontents}

\newcommand{\gr}{$\gamma$-ray}
\newcommand{\grs}{$\gamma$-rays}
\newcommand{\jfactor}{$J$-factor}
\newcommand{\jfactors}{$J$-factors}

\newcommand{\beq}{\begin{equation}}
\newcommand{\eeq}{\end{equation}}

\newcommand{\lcdm}{{\ifmmode \Lambda{\rm CDM} \else $\Lambda{\rm CDM}$\fi}}
\newcommand{\Msol}{M_\odot}
\newcommand{\Msub}{M_{sub}}

\newcommand{\Rsol}{r_\odot}

\definecolor{lightgray}{gray}{0.92}

\title{Search for Galactic dark matter substructures with Cherenkov telescopes}

\ShortTitle{Galactic dark matter substructures with IACT}

\author{\speaker{Moritz H\"utten}
		\\
        Humboldt Universit\"at zu Berlin, Unter den Linden 6, D-10099 Berlin, Germany
        \\ DESY, Platanenallee 6, D-15738 Zeuthen, Germany\\
        E-mail: \email{moritz.huetten@desy.de}}

\author{Gernot Maier\\
        DESY, Platanenallee 6, D-15738 Zeuthen, Germany\\
        E-mail: \email{gernot.maier@desy.de}}

\abstract{
Weakly interacting massive dark matter (DM) particles are expected to
self-annihilate or decay, generating high-energy photons in these processes.
This establishes the possibility for indirect detection of DM by {\gr}
telescopes. For probing the secondary products of DM, accurate knowledge about the
DM density distribution in potential astrophysical targets is crucial.
In this contribution, the prospects for the detection of subhalos in the
Galactic DM halo with present and future imaging atmospheric Cherenkov
telescopes (IACT) are investigated. The source count distribution and angular
power spectra for {\grs} originating from annihilating DM in subhalos are
calculated from N-body simulation results. To study the systematic uncertainties
coming from the modeling of the DM density distribution, parameters describing the
{\gr} yield from subhalos are varied in $16$ benchmark models. We conclude that
Galactic subhalos of annihilating DM are probably too faint to be a promising
target for IACT observations, even with the prospective
Cherenkov Telescope Array (CTA). 
}

\FullConference{The 34th International Cosmic Ray Conference,\\
		30 July- 6 August, 2015\\
		The Hague, The Netherlands}

\begin{document}

\section{Introduction}
\vspace{-0.2cm}
Revealing the nature of dark matter (DM) constitutes one of the most challenging tasks for astrophysics and cosmology. The most promising candidate to explain the numerous observational evidence for DM is a non standard model weakly interacting massive particle (WIMP), which self-annihilates into standard model particles \cite{Bergstrom2009}. 
As the WIMP mass is expected to be in the range of GeV to TeV, it is supposed to
produce secondary {\grs} in the same energy regime. This opens the window for an
indirect detection of DM by {\gr} telescopes. Various techniques exist for the detection of astrophysical {\grs}. This contribution  is focused on imaging atmospheric Cherenkov telescopes (IACT),  in particular the prospective Cherenkov Telescope Array (CTA) \cite{Acharya2013}.

So far, no clear evidence for particle DM, neither direct nor indirect, has been found. Several targets have been considered for the indirect detection of DM, among
them the Galactic center, neighboring dwarf spheroidal galaxies (dSphG), and
galaxy clusters \cite{Conrad2015}. Extensive numerical simulations in the
framework of $\Lambda$CDM cosmology have shown that DM is also expected to form
density structures  on sub-galactic scales \cite{Springel2008, Diemand2008}.
Such subhalos in our Galaxy may constitute an interesting additional target for indirect DM searches. First, they are expected to lack of any contaminating non-DM {\gr} emission, resulting into a signal-to-noise ratio comparable or better than for the dSphG observations.
Second, as the flux of secondary {\grs} is expected to be proportional to the DM density squared, DM subhalos may cause a comparably high small-scale fluctuation of the isotropic {\gr} background (IGRB), detectable and distinguishable from other sources \cite{Ripken2014}.

Many studies have been performed  to assess the {\gr} fluxes from Galactic DM subhalos,
by using the results of numerical simulations, e.g.,
\cite{Siegal-Gaskins2008,Fornasa2012}. However,  uncertainties about the spatial
component of the signal, the \textit{astrophysical $J$-factor},  still
remain. The resolution of numerical N-body simulations is limited to larger
subhalo masses than relevant for a detectable signal. Therefore, extensive
extrapolations and model assumptions have to be applied over many orders of
magnitude. In this contribution, these model systematic uncertainties are
studied with the new version of the semi-analytic code {\sc clumpy}
\cite{Charbonnier2012,Bonnivard2015a}. Four different N-body simulation fit
results and extrapolations are varied, and an uncertainty band  is derived for
the {\jfactors} and the fluctuations in the IGRB from  DM subhalos in the Milky Way.

\section{$\gamma-$rays from Galactic dark matter subhalos} 
\vspace{-0.2cm}
The {\gr} flux $S$ from annihilating dark matter at redshift $z=0$ can be expressed as
\beq
 S =
 \frac{1}{8\pi} \frac{\langle\sigma v\rangle}{m_{\chi}^{\;\,2}}\,
 \int_{E_{\gamma}} 
 \sum_i\, b_i  
 \, \left(\frac{\mathrm{d} N_{\gamma}}{\mathrm{d} E_{\gamma}} \right)_i
 \cdot \int_{\Delta \Omega}\int_{l.o.s.} \rho^2[r(l,\Omega)]\,\mathrm{d}l\,\mathrm{d}\Omega,
\eeq
where $m_{\chi}$ is the dark matter particle's mass, $\langle\sigma v\rangle$
the velocity-averaged annihilation cross section, $b_i$ is the branching ratio for annihilation channel $i$, and $(\mathrm{d}
 N_{\gamma}/\mathrm{d} E_{\gamma})_i$ the final-state {\gr} spectrum in this channel.
The second term, the {\jfactor}, $J=
\int\int\rho^2\,\mathrm{d}l\,\mathrm{d}\Omega$, is the integral of the squared
DM density, $\rho$, over the line of sight, $l$, and $\Delta\Omega = 2\pi
(1-\cos\theta)$, the solid angle over the instrument's aperture radius, $\theta$.
The flux $S$ is proportional to $J$, which is independent of any particle physics model. Therefore, in the following all quantities are expressed in terms of $J$ instead of $S$.

The mean density $\overline{\varrho}$ of the total Galactic DM host halo is modeled by an Einasto profile,
\beq
\overline{\varrho}(r) = \overline{\varrho}_s \exp \left(-\frac{2}{\alpha_E}\left[ \left(\frac{r}{r_s}\right)^{\alpha_E} -1 \right]\right),
 \label{eq:Einasto}
\eeq
on the basis of a best fit to the Aquarius simulations presented in
\cite{Fornasa2012}, with the scale radius $r_s = 15.14\,\rm{kpc}$ and $\alpha_E
= 0.17$. The parameter $1/\alpha_E$ is a measure of the steepness of the density profile close
to $r=0$, with $r$ the distance from the halo center.  Motivated by recent astronomical measurements
\cite{Read2014}, the halo is normalized by a local dark matter density of $0.4\,\rm{GeV\,cm}^{-3}$ at $\Rsol = 8.0\,\rm{kpc}$. For a virial radius of
$260\,\rm{kpc}$ of the host halo, this results in a total halo mass of
$1.1\cdot 10^{12}\,\Msol$. This mass is in agreement with
observations of stellar tracers \cite{Read2014} 
 and is used in the following.

The simulations indicate that the distribution of subhalos in
the host halo follows an anti-biased distribution  with respect to the unclustered,
smooth DM distribution \cite{Springel2008, Diemand2008}. This implies that most
of the clustered mass is located in the outer parts of the host halo. For the model
variation, two parametrizations of the substructure distribution $\overline{\varrho}_{sub}$
within the host halo are compared: The Aquarius-simulation \cite{Springel2008}
suggests an Einasto profile for the subclustered mass with $\alpha_E = 0.678$
and $r_s= 199\,\rm{kpc}$ (model \rm{EINASTO}). The Via-Lactea II subhalos
\cite{Diemand2008} are described by a slightly less anti-biased distribution,
modeled by the  number distribution $N_{sub}(<r)$ given in \cite{Madau2008} (model \rm{MADAU}). All subhalos are described by an Einasto profile with the same $\alpha_E$ as for the
total halo.

The presence of subhalos with masses in the range $[10^{-6}\,\Msol,\, 10^{10}\,\Msol]$ is assumed \cite{Sanchez-Conde2014}\footnote{ Subhalos are drawn by {\sc clumpy} down to a user-defined accuracy in terms of the {\jfactor} contribution from subhalos compared to the background emission from the unclustered ``smooth'' DM in the halo. For all models, all subhalos yielding $J\gtrsim 10^{16}\,{\rm GeV^2\, cm^{-5}}$ are correctly simulated in this study.
}. 
The simulations \cite{Springel2008, Diemand2008} indicate  a power-law function $\mathrm{d}N_{sub}/\mathrm{d}\Msub \sim \Msub^{\;-\alpha_m}$ of the subhalo distribution, with an index $1.9\lesssim \alpha_m \lesssim 2.0$ for the resolved subhalos with masses $\Msub\gtrsim 10^{5}\,\Msol$. The extrapolation of $\mathrm{d}N_{sub}/\mathrm{d}\Msub$ down to the lowest subhalo masses is very sensitive to $\alpha_m$.
Therefore, the results for the two indices $\alpha_m = 1.9$ and $\alpha_m = 2.0$
are compared. A fixed number of $150$ massive dSphG-like subhalos in the
mass range $[10^{8}\,\Msol, 10^{10}\,\Msol]$ is supposed, in agreement with~\cite{Springel2008}.
   
An important property of a subhalo is the concentration $c(\Msub) =
R_{\Delta}/r_{s,\,sub}$, which connects its scale radius $r_{s,\,sub}$ with its virial size
$R_{\Delta}$ as a function of its mass $\Msub$. As the annihilation {\gr} emission is proportional to the density
squared, a higher concentration significantly boosts the brightness of a 
subhalo. The rather optimistic (highly concentrated) model from
\cite{Bullock2001} (model \rm{BULLOCK}) is compared to the conservative one (flatter subhalo concentrations) from \cite{Sanchez-Conde2014} (model \rm{SANCHEZ}). 
Finally, the authors of \cite{Bullock2001} find a significant log-norm
scattering $\sigma_c=\Delta(\log c)$ around the mean concentrations $\bar{c}(\Msub)$, which also results into a net boost of the signal. Their result for hosted subhalos, $\sigma_c=0.24$, is compared to a more  conservative assumption of $\sigma_c=0.14$.

From the two values for each of the four varied quantities, $16$ benchmark models
are created, listed in Tab.~\ref{tab:model-parameters}. Model $9$ is chosen to
be the reference model with an average flux yield.

\begin{table}[!h]
\resizebox{\textwidth}{!}{%
  \begin{tabular}{| l | c | c | c | c | c | c | c | c | c | c | c | c | c | c | c | c |}
    \hline
    Model & 1 & 2 & 3 & 4 & 5 & 6 & 7 & 8 & \cellcolor[gray]{0.8} 9 & 10 & 11 & 12 & 13 & 14 & 15 & 16\\ \hline
    $\overline{\varrho}_{sub}$ & {\rm M} & {\rm E} & {\rm M} & {\rm E} & {\rm M} & {\rm E} & {\rm M} & {\rm E} & \cellcolor[gray]{0.8} {\rm M} & {\rm E} & {\rm M} & {\rm E} & {\rm M} & {\rm E} & {\rm M} & {\rm E} \\ 
    $\alpha_m$ & 1.9 & 1.9 & 2.0 & 2.0 & 1.9 & 1.9 & 2.0 & 2.0 & \cellcolor[gray]{0.8} 1.9 & 1.9 & 2.0 & 2.0 & 1.9 & 1.9 & 2.0 & 2.0 \\ 
    $c(\Msub)$ & {\rm S} & {\rm S} & {\rm S} & {\rm S} & {\rm B} & {\rm B} & {\rm B} & {\rm B} & \cellcolor[gray]{0.8} {\rm S} & {\rm S} & {\rm S} & {\rm S} & {\rm B} & {\rm B} & {\rm B} & {\rm B} \\ 
    $\sigma_c$ & 0.14 & 0.14& 0.14 & 0.14 & 0.14 & 0.14 & 0.14 & 0.14 & \cellcolor[gray]{0.8} 0.24 & 0.24 & 0.24 & 0.24 & 0.24 & 0.24 & 0.24 & 0.24 \\ 
    \hline \hline
     $\alpha_J$ & 2.05 & 2.06 & 2.15 & 2.15 & 2.03 & 2.02 & 2.12 & 2.13 & \cellcolor[gray]{0.8} 2.04 & 2.05 & 2.14 & 2.15 & 2.03 & 2.04 & 2.09 & 2.12 \\ 
    \hline
  \end{tabular}
  }
  \caption{Parameters for the $16$ models investigated in this study. The reference model is marked in gray. {\rm S}/{\rm B} stands for the {\rm SANCHEZ}/{\rm BULLOCK} $c(\Msub)$ parametrization, {\rm E}/{\rm M}: $\overline{\varrho}_{sub}$ parametrization by an \rm{EINASTO}/\rm{MADAU} profile. Bottom row: Fit values for the exponent $\alpha_J$ of a power law fit for $\mathrm{d}N_{sub}/\mathrm{d}J\sim J^{\alpha_J}$ in the regime $J<10^{17}\;{\rm GeV^2\, cm^{-5}}$. The fitting uncertainty is $\sigma_{\alpha_J}\lesssim 5\cdot 10^{-2}$ for all models.}
\label{tab:model-parameters}
\end{table}

\section{Setup of the simulations}
\vspace{-0.2cm}

The Galactic DM halo is simulated for a circular field of view (FOV) centered at
the Galactic pole $b=90^{\circ}$ with diameter $d= 120^{\circ}$. This
FOV covers a fraction $f_{sky}=25\%$ of the sky. The choice of the FOV is
motivated by the projected extragalactic sky survey achievable with CTA \cite{Dubus2013}, an area where it can be searched also for Galactic DM subhalos.
To take into account the statistical variation of the observed sky in terms of
 cosmic variance, each simulation is rerun $10.000$ times. In the same way, the search for DM subhalos
within a single FOV of $d= 10^{\circ}$ is investigated, using a separate simulation.

\section{Results}
\vspace{-0.2cm}

\begin{figure}[!t]
\centering
\includegraphics[width=14cm]{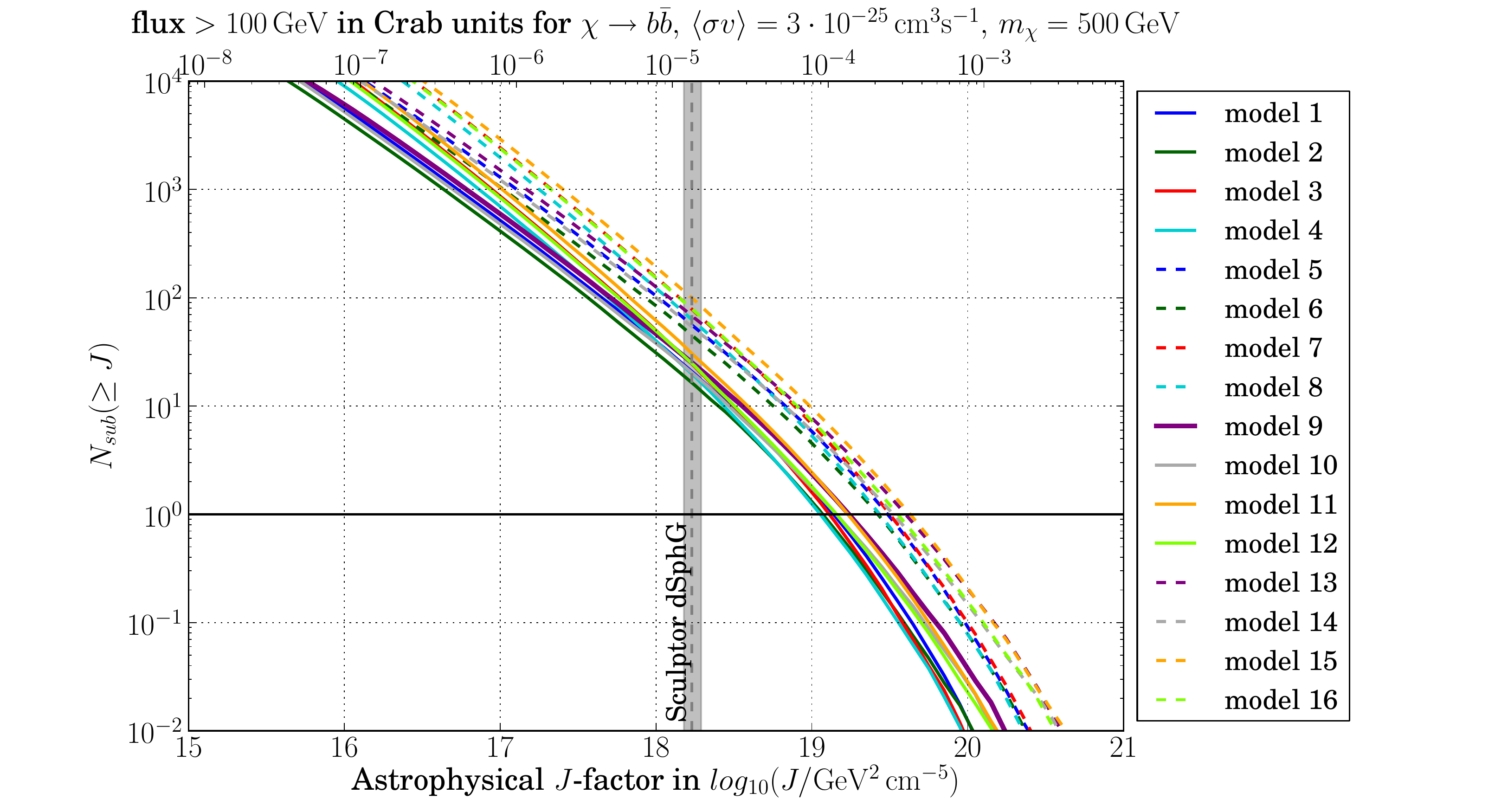}
\caption{Integrated source count distribution, $N_{sub}$, for all 16 benchmark
models on $f_{sky}=0.25$ and for
$\theta_1 = 0.1^{\circ}$. The dependence of $N_{sub}$ on the {\jfactor} is shown on the lower abscissa. On the upper abscissa, the dependence of $N_{sub}$ on the
flux is shown, for the assumed particle physics model.
The continuous/dashed lines distinguish the {\rm SANCHEZ}/{\rm BULLOCK}
concentration model. The corresponding {\jfactor} from Sculptor dSphG for
$\theta_1 = 0.1^{\circ}$ \cite{Geringer-Sameth2015} is plotted for comparison.}
\label{fig:PopStudy-all}
\vspace{0.5cm}
\begin{minipage}[h]{.485\textwidth}
  \begin{flushleft}
  \vspace{0.03cm}
  \includegraphics[width=0.97\textwidth]{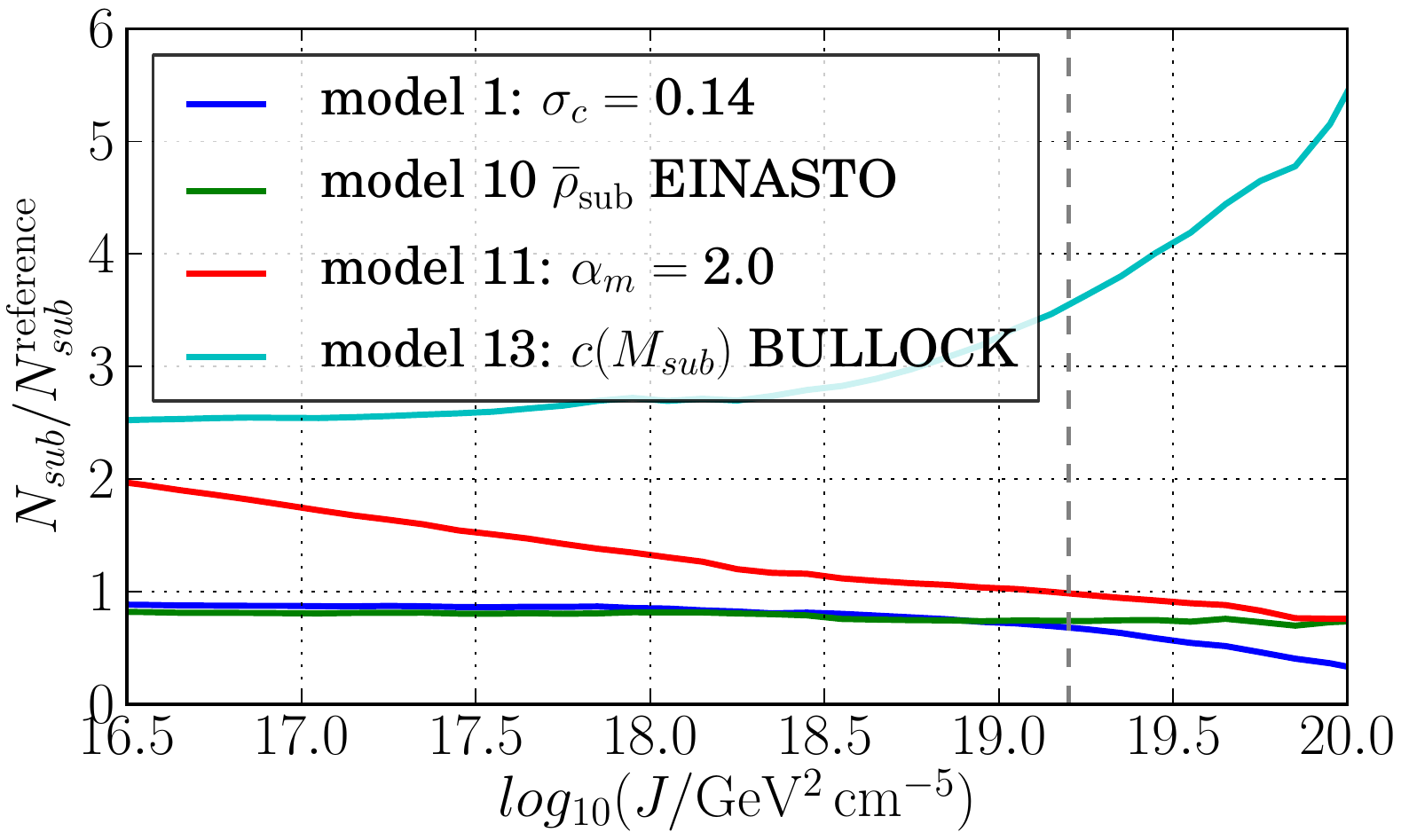}
  \caption{$N_{sub}$ of the four models that differ only by the depicted parameter from  model 9 ($\sigma_c=~0.24$, $\overline{\varrho}_{sub}$ MADAU, $\alpha_m=1.9$, $c(\Msub)$ SANCHEZ), normalized to $N_{sub}$ of model 9.
The dashed line marks the $J$ of the brightest subhalo expected ($N_{sub}(\geq~J)~=~1$) for  model 9.
}
  \label{fig:PopStudy-comparisons-all}
  \end{flushleft}
\end{minipage}
\hspace{.01\textwidth}
\begin{minipage}[h]{.485\textwidth}
  \begin{flushright}
  \includegraphics[width=\textwidth]{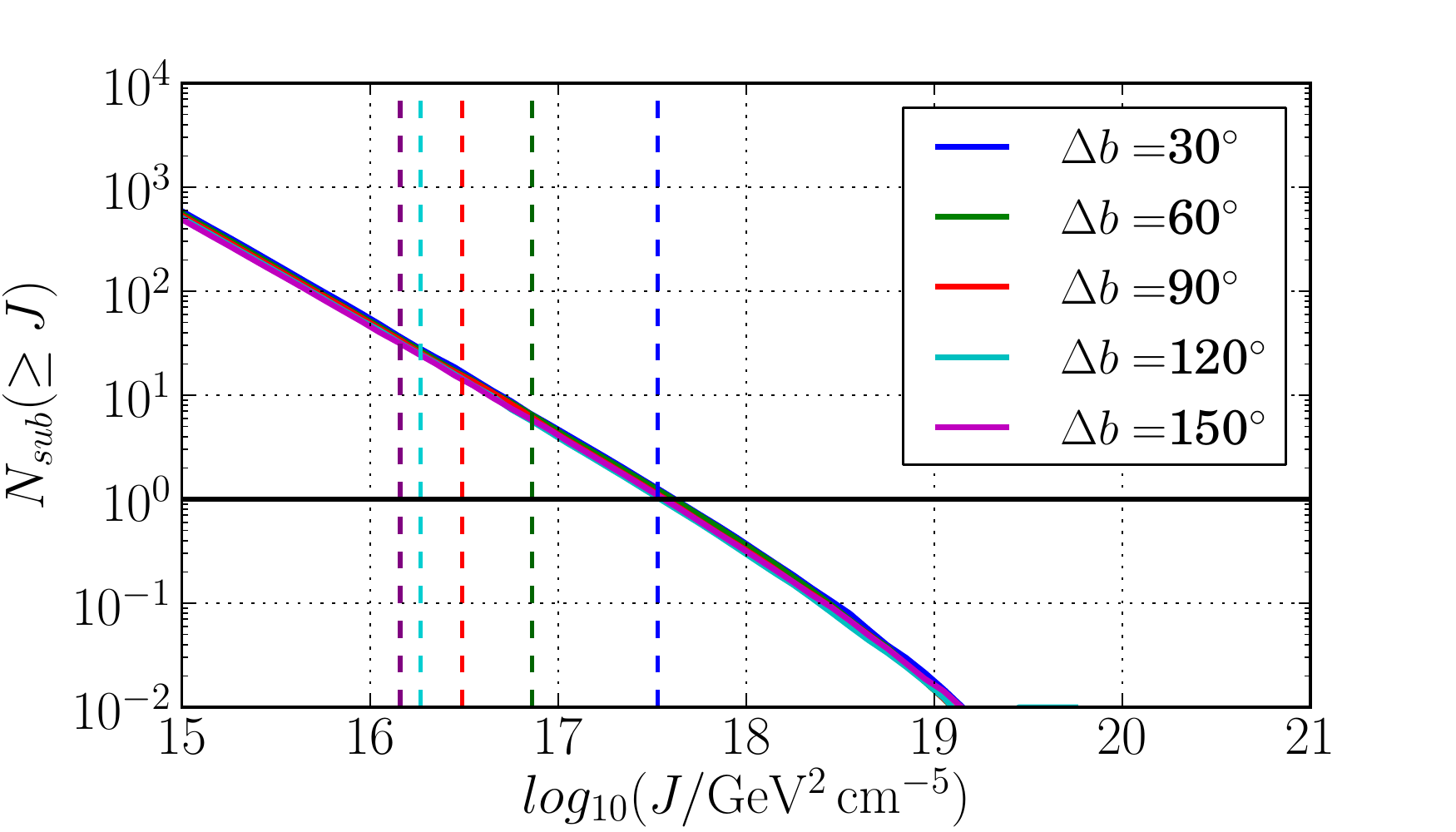}
  \caption{$N_{sub}$ for  model 9 and a FOV diameter of $d=10^{\circ}$ at different distances $\Delta b$ from the Galactic center. The maximum difference between all solid curves is $<10\%$. 
  The dashed lines denote the mean emission from the smooth galactic DM
  component for each FOV, and $\theta_1 = 0.1^{\circ}$.}
  \label{fig:PopStudy-1FOV}
  \end{flushright}
\end{minipage}
\vspace{-0.3cm}
\end{figure}

In Fig.~\ref{fig:PopStudy-all}, the average integrated \textit{source count
distribution} is shown, defined as
\beq
N_{sub}(\geq J) = \int\limits_J^{\infty} \frac{\mathrm{d}N_{sub}}{\mathrm{d}J'}\,\mathrm{d}J'.
\eeq
Here, $N_{sub}(\geq J)$ represents the number of subhalos on the chosen
patch of the sky brighter than a given {\jfactor}.
The DM subhalo source count distribution $\mathrm{d}N_{sub}/\mathrm{d}J$  can be described by a power-law according to
$\mathrm{d}N_{sub}/\mathrm{d}J\sim J^{-\alpha_J}$ in the
regime $J<10^{17}\;{\rm GeV^2\, cm^{-5}}$.  In Tab.~\ref{tab:model-parameters}, the fit results of the exponent $\alpha_J$  are listed for each of the $16$
models. These results indicate that  the $\mathrm{d}N_{sub}/\mathrm{d}J$
distribution is dominantly determined by the mass distribution
$\mathrm{d}N_{sub}/\mathrm{d}\Msub\sim \Msub^{\;-\alpha_m}$. 

\begin{wrapfigure}{r}{0.4\textwidth}
     \includegraphics[width=0.4\textwidth]{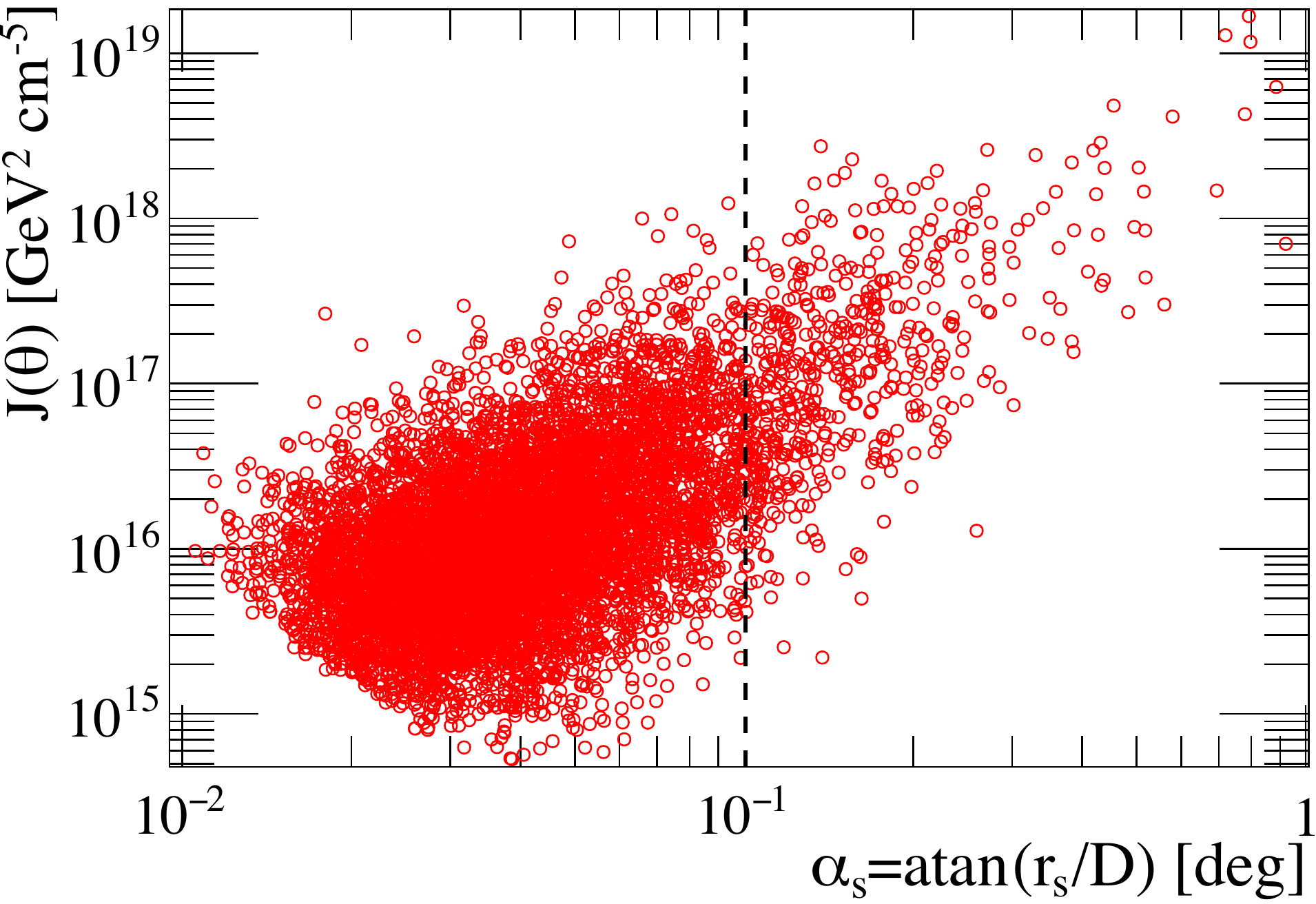}
   \setlength{\abovecaptionskip}{-10pt plus 3pt minus 2pt}
  \caption{Relation between the angular size $\alpha_s$ and the brightness
  $J(\theta_1= 0.1^{\circ})$ of the subhalos for the survey region \& model 9.
  $D$ is the subhalo distance from the observer. The dashed line indicates $\alpha_s=\theta_1$.}
  \label{c_subs_j_alphas}
  \vspace{-15pt}
\end{wrapfigure}

The integration angle, within which the {\jfactor} of each subhalo is calculated, is set to $\theta_1 =
0.1^{\circ}$ to illustrate the power-law behavior of the source count
distribution. In Fig.~\ref{c_subs_j_alphas}, it is shown for model 9 that the largest part of most of the subhalos is encompassed within $\theta_1$. Similar results are obtained for all models. As the annihilation {\gr} emission is strongly
peaked at the subhalo center, only the signal of the most extended subhalos
gets  significantly truncated by $\theta_1$. Fig.~\ref{c_subs_j_alphas}
suggests a correlation between the  apparent size and the brightness of the
subhalos.
Therefore, a cut-off of the power-law distribution at high $J$ is visible in
Fig.~\ref{fig:PopStudy-all}. Continuous emission from the unclustered DM (``smooth halo'') and other background decreases proportional to $\theta^{2}$, faster than the subhalos' flux \cite{Bonnivard2015}. Consequently, a smaller signal region might  
improve the signal-to-noise ratio in a real experiment.

Only a slight correlation
between the subhalo masses and their {\jfactors} has been found for all models (not shown). From
this follows that  distant  massive dSphG-like subhalos and nearby  low-mass  subhalos
similarly account for the highest {\jfactors}.

  For a given DM annihilation channel, cross section and particle mass,  the
 fluxes corresponding to the {\jfactors} are plotted in Fig.~\ref{fig:PopStudy-all}. For the annihilation channel $\chi\chi\rightarrow b\bar{b}$, CTA will be most sensitive to a DM particle mass of $500
 \,\mathrm{GeV} \lesssim m_{\chi} \lesssim 1
 \,\mathrm{TeV}$ \cite{Carr2015}. For this mass range, the annihilation cross section is already constrained to  $\langle \sigma v \rangle \lesssim 3\cdot 10^{-25}\,\mathrm{cm^3\,s^{-1}}$ by observations of the dSphGs by the \textit{Fermi}-LAT instrument~\cite{Ackermann2015}. Thus, in the range of the considered models and a coverage of $f_{sky} = 25\%$, a survey sensitivity to some $10^{-4}$ of the Crab Nebula flux is necessary to discover at least one subhalo above $100\,\mathrm{GeV}$ with IACTs, provided that $\langle \sigma v \rangle \approx 3\cdot 10^{-25}\,\mathrm{cm^3\,s^{-1}}$.

Fig.~\ref{fig:PopStudy-comparisons-all} shows the impact of the varied parameters on $N_{sub}$ compared to the reference model. For the brightest subhalos, the model uncertainty is generally $\lesssim 50 \%$. An exception from this occurs when switching from the \rm{SANCHEZ} to the \rm{BULLOCK} concentration model for $c(\Msub)$. In this case, the number of brightest subhalos is increased by more than a factor of three.

Fig.~\ref{fig:PopStudy-1FOV} shows the simulation results for $N_{sub}$ within a
single CTA-like FOV of $d=10^{\circ}$. Here, $N_{sub}$ is given for different distances $\Delta b$ from
the Galactic center, perpendicular to the Galactic plane. Two conclusions can
be drawn from this. First, Fig.~\ref{fig:PopStudy-1FOV} indicates that at
$\Rsol$, the subhalos appear almost isotropically in the sky. Consequently, the
relation $N_{sub}(\Delta \Omega_1)/N_{sub}(\Delta \Omega_2)\approx \Delta \Omega_1/\Delta
\Omega_2$ holds for arbitrary positions in the sky (and is confirmed by
comparing Fig.~\ref{fig:PopStudy-all} and \ref{fig:PopStudy-1FOV}). However,
pointing close to the Galactic center heavily decreases the contrast between the
subhalos and the  smooth DM halo emission, the latter indicated by the dashed lines.
Second, a sensitivity to {\jfactors} of $J\lesssim 10^{18}\;{\rm GeV^2\, cm^{-5}}$ is needed to resolve at least one subhalo within a given single FOV. Such {\jfactors} are about a factor 10 smaller than those of the most promising dSphGs, where observation times of $\gg 500\,\mathrm{hrs}$ are expected to reach competitive limits $\langle\sigma v\rangle\approx 10^{-25}\,\mathrm{cm^3\,s^{-1}}$ with CTA \cite{Carr2015}. 

\begin{figure}[!h]
\vspace{-0.3cm}
\centering
\includegraphics[width=7.5cm]{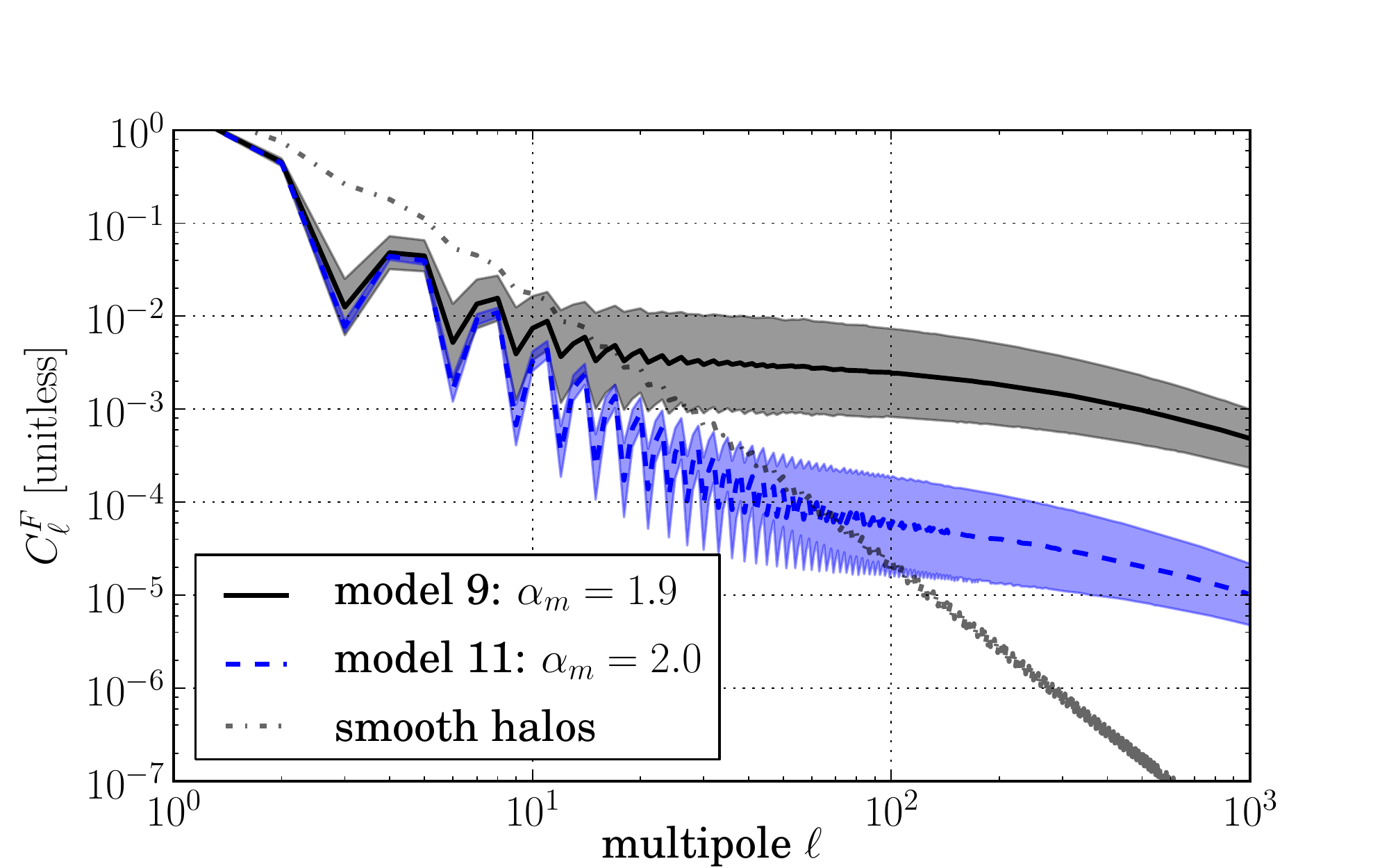}
\includegraphics[width=7.5cm]{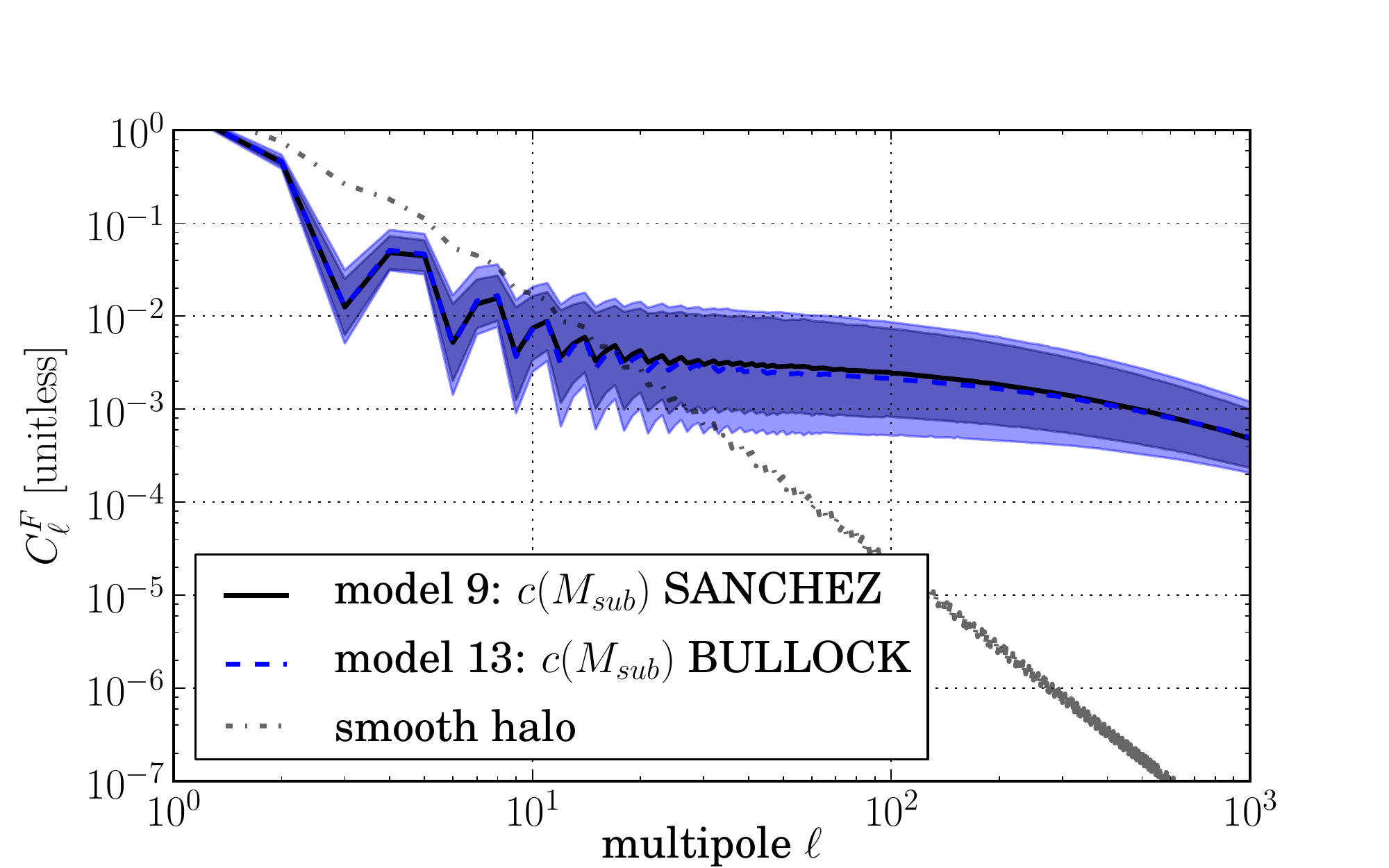}
\caption{Fluctuation APS $C_{\ell}^F$ of the subhalos on $f_{sky}= 0.25$, comparing the
variation of the index $\alpha_m$ (left) and the $c(M_{sub})$ model (right)  against  model 9. The fluctuation APS of the smooth halo on the simulated patch of the sky is plotted for comparison. The
 bands indicate the $1\sigma$ containment ranges around the mean spectrum of the
 simulated galactic halo, assuming a log-normal variation of the
 $C_{\ell}^F$. 
 Note that for the left panel, also the smooth contribution varies between the compared models. However, this difference in $C_{\ell}^F$ between the smooth contributions is smaller than the resolution of the plots and not distinguished here.
}
\label{fig:APS-comparisons}
\end{figure}

In Fig.~\ref{fig:APS-comparisons}, the \textit{fluctuation angular power spectra} (fluctuation APS)
of the subhalos are shown, comparing two of the four varied parameters against
the reference model.  The fluctuation APS is a measure of the relative fluctuation, or \textit{small scale anisotropy}, of a signal, which can be possibly detected above a much larger background holding less fluctuation. The fluctuation APS $C_{\ell}^F$ of a map $I(\vartheta, \varphi)$, defined on a patch $f_{sky}$ of the sphere ($I\equiv 0$ elsewhere) is given by
\beq
C_{\ell}^F = \frac{1}{f_{sky} \; \overline{I}^2\;(2\ell + 1)} \sum\limits_{m= -\ell}^{+\ell} |a_{\ell m}|^2 \;,
\label{eq:fluctuation-aps}
\eeq
with the spherical coordinates $\vartheta$ and $\varphi$, and $\overline{I}$
the mean value of $I$ on the defined domain. The  $a_{lm}$ are the coefficients
of the map decomposed into spherical harmonics $Y_{lm}$,
\beq
I(\vartheta, \varphi) = \sum\limits_{\ell= 0}^{\ell_{max}} \sum\limits_{m= -\ell}^{+\ell} a_{\ell m}\, Y_{\ell m}(\vartheta, \varphi).
\label{eq:multipoledecomp}
\eeq

For Fig.~\ref{fig:APS-comparisons}, an integration angle $\theta_2 =
0.015^{\circ}$ is chosen, which corresponds to half the value of the maximum resolution projected for
CTA at energies above $1\,\mathrm{TeV}$ \cite{Dubus2013}.

 For a FOV with diameter $d\lesssim 10^{\circ}$ and an  angular resolution
 of $\sigma_{PSF}\gtrsim 0.03^{\circ}$ \cite{Dubus2013}, CTA could be sensitive
 to multipoles $100 \lesssim  \ell  \lesssim 1000$ \cite{Ripken2014}. As shown
 in Fig.~\ref{fig:APS-comparisons}, in this regime, the fluctuation APS remains
 largely constant for all models, with only a mild lowering of the amplitudes
 $C_{\ell}^F$ towards higher multipoles.  Only in the case of the variation of
 $\alpha_m$ from $1.9$ to $2.0$, a drastic change for the fluctuation APS
 results. This  is shown in  Fig.~\ref{fig:APS-comparisons} (left).  For
 $\alpha_m = 2.0$, a substantially higher amount of total mass is bound into
 subhalos.  Although this results into a boost of the overall signal, the
 effective fluctuation heavily decreases because of the large background of the
 low-mass subhalos. For all other cases, the fluctuations of the subhalos
 remain largely the same. This applies even for the significant signal boost
 for the {\rm BULLOCK} case, which is shown in Fig.~\ref{fig:APS-comparisons} (right).

\section{Conclusions}
\vspace{-0.2cm}
Different models of the abundance and properties of Galactic DM subhalos have
been compared. Within the varied parameters, the source count distribution
$\mathrm{d}N_{sub}/\mathrm{d}J$ of these subhalos has been constrained to an
uncertainty band of one order of magnitude. The modeling of the inner
concentration $c(\Msub)$ causes the biggest systematic uncertainty in the number of detectable subhalos. The subhalos appear nearly isotropically distributed in the sky, allowing  a telescope to point as far as possible away from  the Galactic plane to reduce the background from the smooth DM halo emission and diffuse astrophysical foregrounds.

However, for all benchmark models, Galactic subhalos are probably too faint to
be a promising target for current and
next-generation IACTs, provided the projected available observation times. Still for a sensitivity  to a fraction of $10^{-4}$ of the
Crab Nebula flux above $100 \,\mathrm{GeV}$, a detection of
individual subhalos might be possible only for cross sections $\langle \sigma v \rangle \gtrsim 10^{-25}\,\mathrm{cm^3\,s^{-1}}$,  which are already in tension with recent measurements.  Pointed observations of the Galactic center and the dSphGs with CTA promise competitive limits on the annihilation cross section, probably requiring much less observation time.

At the same time, it has been shown that a  source count distribution with
$\alpha_J\lesssim 2.1$ results in a  fluctuation APS up to the order of $10^{-3}
\lesssim C_{\ell}^F \lesssim 10^{-2}$ in the multipole range of interest of
Cherenkov telescopes. In principle, such a comparably high fluctuation might  distinguish dark matter from other, more
dominant unresolved source classes. Possibly, even more
competitive limits on the annihilation cross section can be obtained 
than by the search for individual subhalos~\cite{Ripken2014}. However, given the low intensity of the overall signal from Galactic DM subhalos, these limits are most likely still not competitive with those obtainable from (combined) pointed observations.

We did not assess the extragalactic contribution to DM induced fluctuations of the IGRB. However, the results of \cite{Fornasa2012} indicate that the extragalactic emission is expected to be similar to the Galactic signal in magnitude and in angular scales, suggesting a similar sensitivity to IGRB fluctuations as for the Galactic subhalo case.
Still, it should be emphasized that all indirect DM targets suffer from major
systematic uncertainties. With this in mind, the search for  fluctuations in the IGRB should not be neglected as a complementary strategy to unveil the nature of DM.

\section*{Acknowledgments}
\vspace{-0.3cm}
This work is supported by the Research Training Group 1504 of the German
Research Foundation (DFG) and a travel grant by the German Academic Exchange
Service (DAAD). Some of the results in this contribution have been derived using the {\sc HEALPix} package (G\'orski et al., 2005).

\bibliographystyle{JHEP}
\bibliography{ICRC2015-Proceedings_arxiv}

\end{document}